\documentstyle[psfig,preprint,aps]{revtex}
\begin{document}
\tolerance 10000

\draft

\title{Magnetic and Charge Correlations of the 2-dimensional
$t-t'-U$ Hubbard model}

\author{Raymond Fr\'esard$^{1}$\thanks{E-mail:Raymond.Fresard@iph.unine.ch
} and Walter Zimmermann$^{2}$
\thanks{E-mail: walter@tkm.physik.uni-karlsruhe.de}}
\address{
$^{1}$ Institut de Physique, Universit\'e de Neuch\^atel, A-L Breguet 1,
2000 Neuch\^atel, Switzerland\\
$^2$ Institut f\"ur Theorie der Kondensierten Materie, Physikhochhaus, 
Postfach 6980,\\
Universit\"at Karlsruhe, 76128 Karlsruhe, Federal Republic of Germany}

\date{Received}
\maketitle
\widetext

\vspace*{-1.0truecm}

\begin{abstract}
\begin{center}
\parbox{14cm}{Using  a spin-rotation invariant six-slave boson
representation, we study 
the square lattice Hubbard model with nearest-neighbor hopping $t$ and
next-nearest neighbor 
hopping $t'$. We discuss the influence of $t'$ on the charge and
magnetic properties. In the hole-doped domain, we find that a negative
$t'$ strongly favors
itinerant ferromagnetism over any incommensurate phase, especially in
the strong coupling regime.
For positive $t'$ magnetic fluctuations are suppressed. A
tight connection between frustrated charge dynamics and large magnetic
fluctuations is pointed out. A clear tendency towards striped charge
ordering is found in the regime of large positive $t'$.
} 
\end{center}
\end{abstract}

\pacs{
\hspace{1.9cm}
PACS numbers: 71.10.Fd 75.40.Cx 74.25.Ha }
\section{Introduction} 
The spin rotation invariant slave boson representation is applied to
the $t-t'-U$-model. This model is expected to be relevant to the
physics of high-temperature superconductors, since
it includes a reasonable description of their band structure. It is a
good candidate for
the description of itinerant ferromagnetism too. Both
behaviors are expected to occur in different regions of the phase
diagram. Indeed it can be thought of as consisting of three characteristic
regions, depending on whether the magnetic fluctuations are strong or
not, and whether they are ferromagnetic or anti-ferromagnetic. The
size of these regions is tuned by $t'$. In the non-interacting
limit the role of $t'$ is to shift the van Hove singularity that lies
in the middle of the band for $t'=0$ to the lower band edge for
$t'=-t/2$ or to the upper band edge for $t'=t/2$. The extension of
this physics to the weak coupling regime has been extensively studied
by Lin and Hirsch \cite{Lin}, and B\'enard {\it et al.} \cite{Benard}, and
Lavagna and Stemman \cite{Stemman}. They found that, for large
negative $t'$,
the physics is dominated by strong ferromagnetic fluctuations in the
low density domain, and by strong antiferromagnetic fluctuations in
the vicinity of half-filling. Quantum Monte Carlo (QMC) simulations
have been performed too. In particular Veilleux {\it et al.} \cite{Veilleux}
confirmed this behavior, and thus put it on a stronger basis. They
also established that the static and uniform magnetic susceptibility
goes over a maximum when the system is doped off half-filling. Recently
Hlubina {\it et al.} \cite{Hlubina} studied the same model at densities
corresponding to the van Hove singularity, and found that the system
is an itinerant ferromagnet for large negative $t'$, and an itinerant
antiferromagnet for small negative $t'$. Unfortunately these techniques can
only be applied in the weak to intermediate coupling regime, because
of the minus sign problem for QMC simulations, and because the RPA is
intrinsically a weak coupling approach. 
For strong coupling one usually resorts to variational
methods \cite{Pieri,Muller}, (for a recent discussion see
\cite{Vollhardt}). In order to cover the entire
parameter range it is tempting to apply the Kotliar and Ruckenstein
slave boson approach \cite{Kotliar} . It not only proved to yield
ground state energies
very close to the exact ones, but very realistic values for the
structure factors too \cite{Zimmermann}. Until now little
attention has been paid to the charge structure factor.
The aim of this brief report is two-fold. First we calculate the
$q$-dependent magnetic susceptibility in order to determine in which
domain of the phase diagram the antiferromagnetic, ferromagnetic and
incommensurate fluctuations dominate, both in the intermediate and
strong coupling regime. Second we calculate the charge structure
factor. We then show that strong magnetic fluctuations are
systematically accompanied with a clear reduction of the charge
structure factor, i.e. by a frustration of the charge dynamics. In
this model this happens for negative $t'$, in the hole doped
region. For positive $t'$, the charge structure factor is enhanced and
the magnetic fluctuations suppressed. We obtained this result using the
spin-rotation invariant (SRI) slave boson representation of the
Hubbard model \cite{Hirschfeld,Fresard}, and the expression
for the spin and charge dynamical susceptibilities we recently
derived and applied to the Hubbard model \cite{Zimmermann}. We note
that this representation has been recently revisited by Ziegler {\it
et al.} \cite{Ziegler}. Our expressions for the susceptibilities remain
unchanged by their considerations.

\section{Formalism}
In this work we calculate the spin and charge dynamical
susceptibilities of the two dimensional $t-t'-U$ model. The
Hamiltonian reads:
\begin{equation}
H=-\sum_{i,j,\sigma}t_{i,j}c_{i\sigma}^{+}c_{j\sigma}+
U\sum_{i}n_{i\uparrow}n_{i\downarrow}
\end{equation}
We consider the case where the hopping integral is $t_{i,j}=t$ for
nearest neighbors, $t_{i,j}=t'$ for next-nearest neighbors and
$t_{i,j}=0$ otherwise.
To this aim we apply the spin-rotation invariant (SRI) slave 
boson formulation of the Hubbard model  \cite{Hirschfeld,Fresard} to the
one-loop calculation of the susceptibilities that we applied to the
Hubbard model. In this framework the dynamical spin susceptibility is
given by
\begin{equation} \label{chis}
\chi_{s}({\vec k},\omega)=\frac{\chi_{0}({\vec k},\omega)}
{1+A_{\vec k} \chi_{0}({\vec k},\omega)+A_{1}\chi_{1}({\vec k},\omega)+
A_{2}[\chi_{1}^{2}({\vec k},\omega)-\chi_{0}({\vec
k},\omega)\chi_{2}({\vec k},\omega)]}\quad , 
\end{equation}
where
\begin{eqnarray}\label{blob}
\chi_{n}({\vec k},\omega)  & = & -\sum_{{\vec p}, i\omega_n, \sigma}
(t_{\vec p}+t_{\vec p +\vec k})^{n}
G_{0\sigma}({\vec p}, i\omega_n)G_{0\sigma}({\vec p} + {\vec k}, 
\omega+i\omega_n) \quad (n=0,1,2)
\end{eqnarray}
In the low frequency regime Eq. (\ref{chis}) has an RPA form, to which
it reduces in the weak coupling limit. It
nevertheless differs from it in two important respects. First the
effective interaction $A_{{\vec k}}$ does not grow indefinitely as $U$ grows,
but saturates at a fraction of the average kinetic energy in the strong
coupling regime. Second it is
$k$-dependent. Therefore if an magnetic instability of the
paramagnetic phase at a given density develops towards an
incommensurate phase, characterized by a wave vector $\vec q$, this
wave vector will be different from the wave vector $\vec p$ at which
$\chi_0(\vec p,0)$ reaches its maximum. Thus at a given density, the
wave-vector $\vec q$ characterizing the phase towards which the
paramagnetic phase can be unstable to, depends on the interaction
strength, in contrast to the ordinary RPA. It numerically turns out that the
$\vec k$-dependence of $A_{{\vec k}}$ is enhanced by increasing
the interaction strength. $A_{{\vec k}}$ is typically largest for
$k=0$, and such is
$\chi_1(\vec k)$. The contribution involving $A_2$ is smallest, and
has little influence on the magnetic properties. The numerous 
undefined symbols
in Eqs. (\ref{chis}-\ref{blob}) can be gathered from
Ref. \cite{Zimmermann}, except for a misprint there: the third line of
Eq. (A8) should read:
\begin{equation}
\frac{\partial^2 z }{\partial d'^2} = \frac{2 \sqrt{2} p_0
\eta}{1+\delta} \left( 2d + x + \frac{6 x d^2}{1+\delta}\right) \quad .
\end{equation}

\section{Results}
\begin{figure}
\narrowtext
\centerline{\psfig{figure=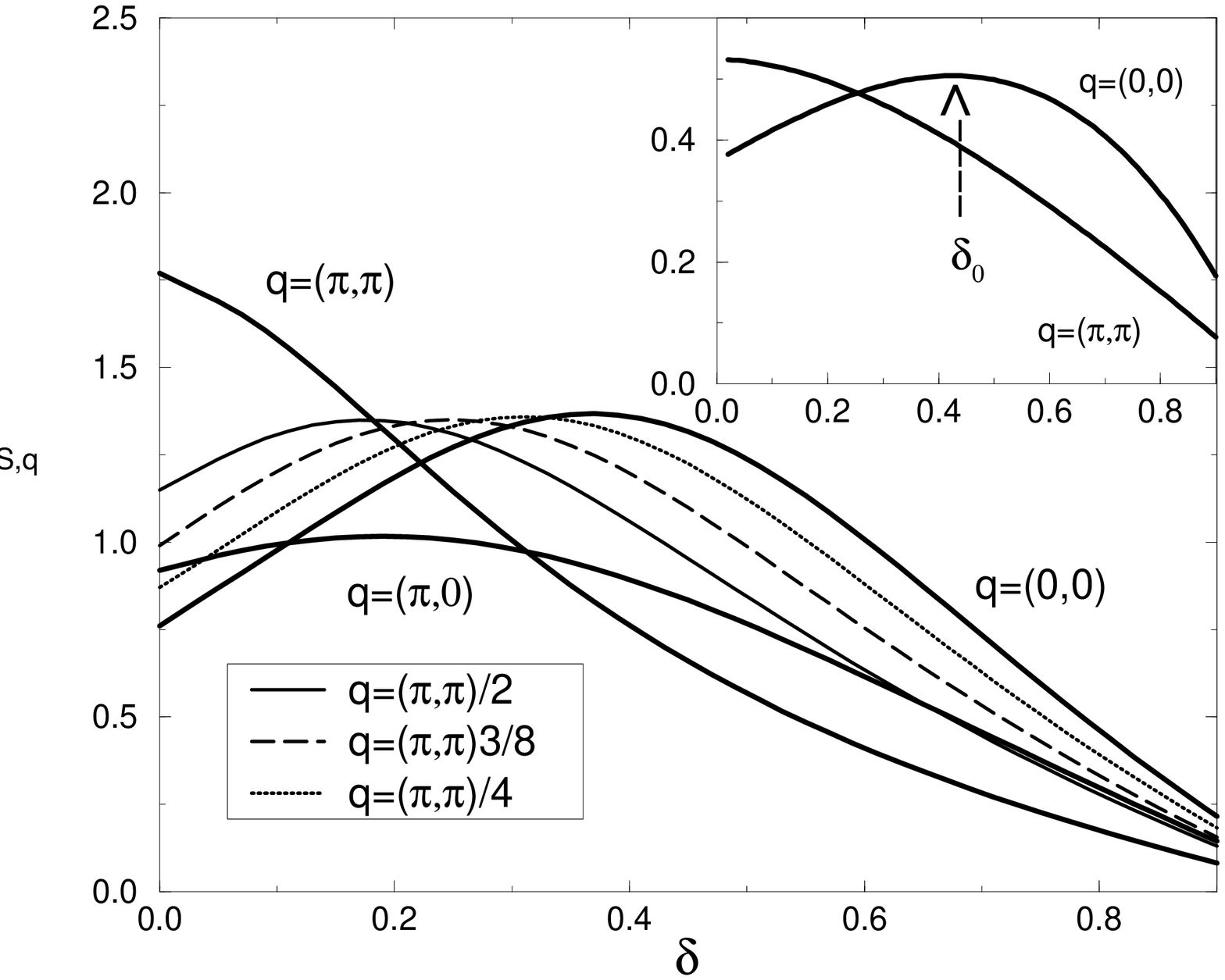,height=16cm,width=16cm}}
\caption{\label{fig1} Doping-dependence of the magnetic susceptibility
for $U=4t$, $\beta=2$ and $t'=-0.47t$ for a set of wave-vectors lying
along the three main directions of the Brillouin zone. Inset: 
Doping-dependence of the magnetic susceptibility for $U=0$ for the 
anti-ferro- and ferromagnetic wave-vector.} 
\end{figure}

\begin{figure}
\narrowtext
\centerline{\psfig{figure=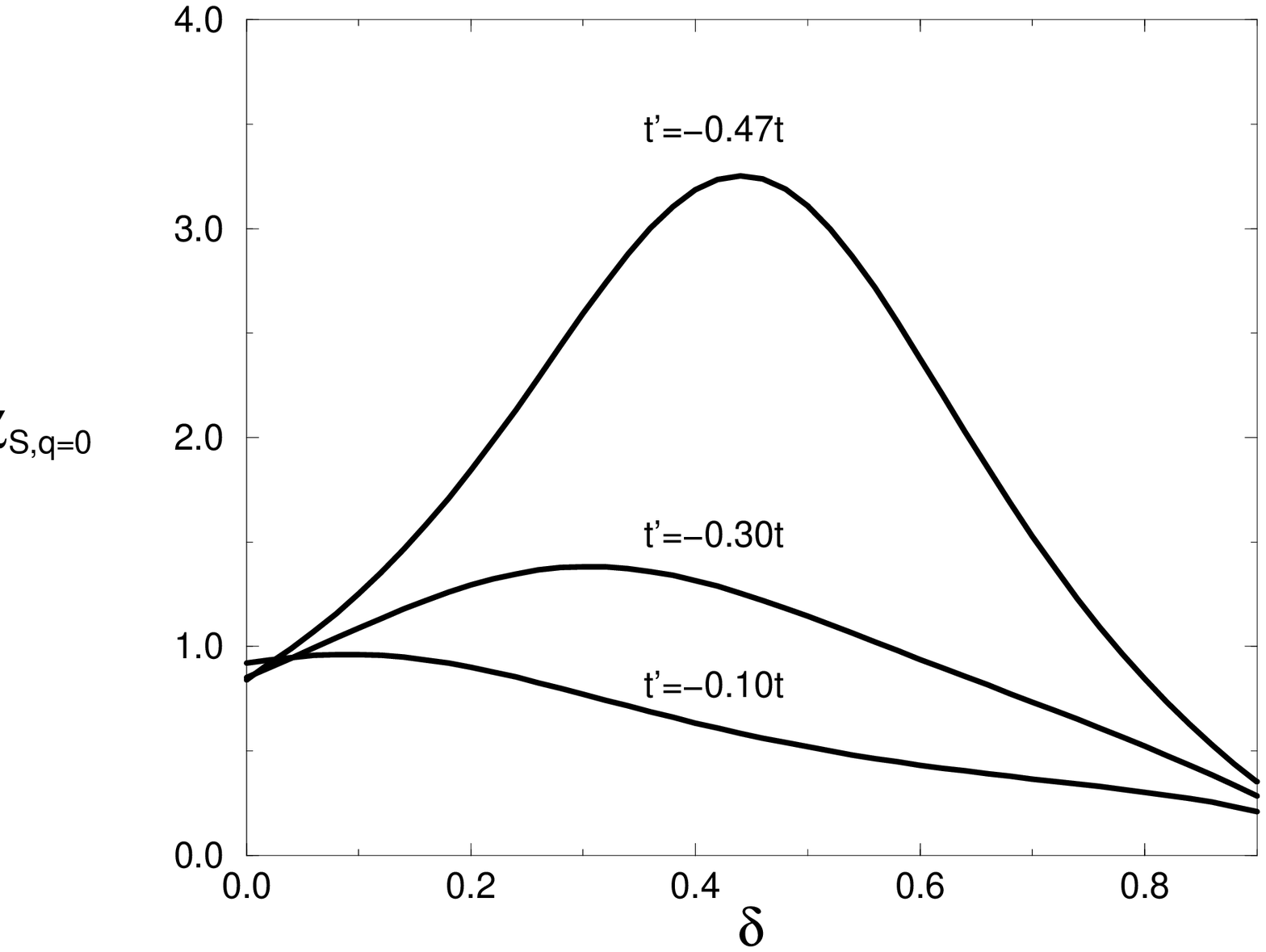,height=16cm,width=16cm}}
\caption{\label{fig2} Doping-dependence of the static and uniform
magnetic susceptibility
for $U=4t$, $\beta=3$ and $t'=-0.1t$, $t'=-0.3t$ and $t'=-0.47t$.} 
\end{figure}

\begin{figure}
\narrowtext
\centerline{\psfig{figure=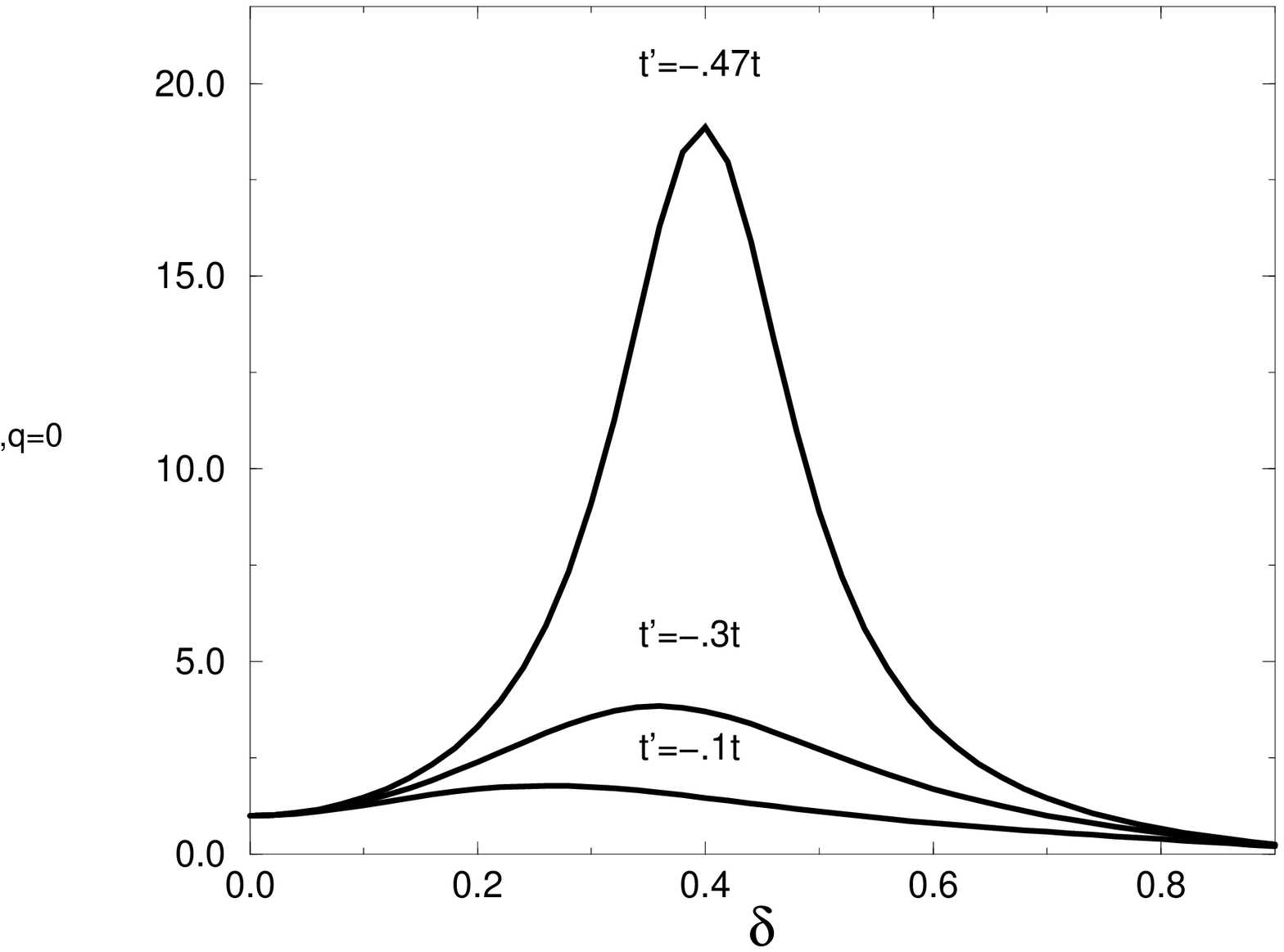,height=16cm,width=16cm}}
\caption{\label{fig3} Doping-dependence of the static and uniform
magnetic susceptibility
for $U=20t$, $\beta=2$ and $t'=-0.1t$, $t'=-0.3t$ and $t'=-0.47t$.} 
\end{figure}

\begin{figure}
\narrowtext
\centerline{\psfig{figure=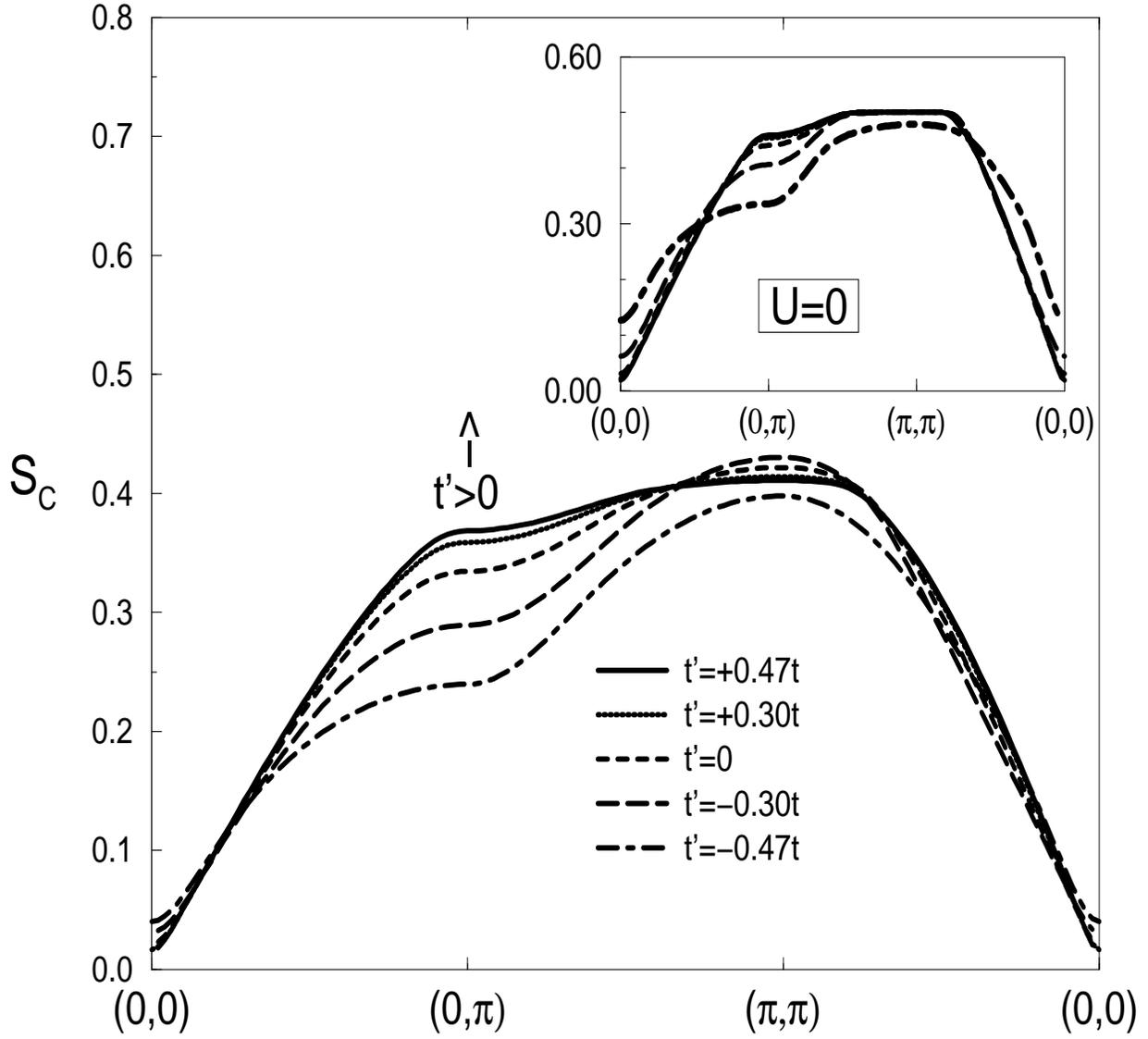,height=16cm,width=16cm}}
\caption{\label{fig4} Charge structure factor of the t-t'-U model at
quarter filling, $U=4t$ and $\beta=8$ and for several values of t'.}  
\end{figure}

\begin{figure}
\narrowtext
\centerline{\psfig{figure=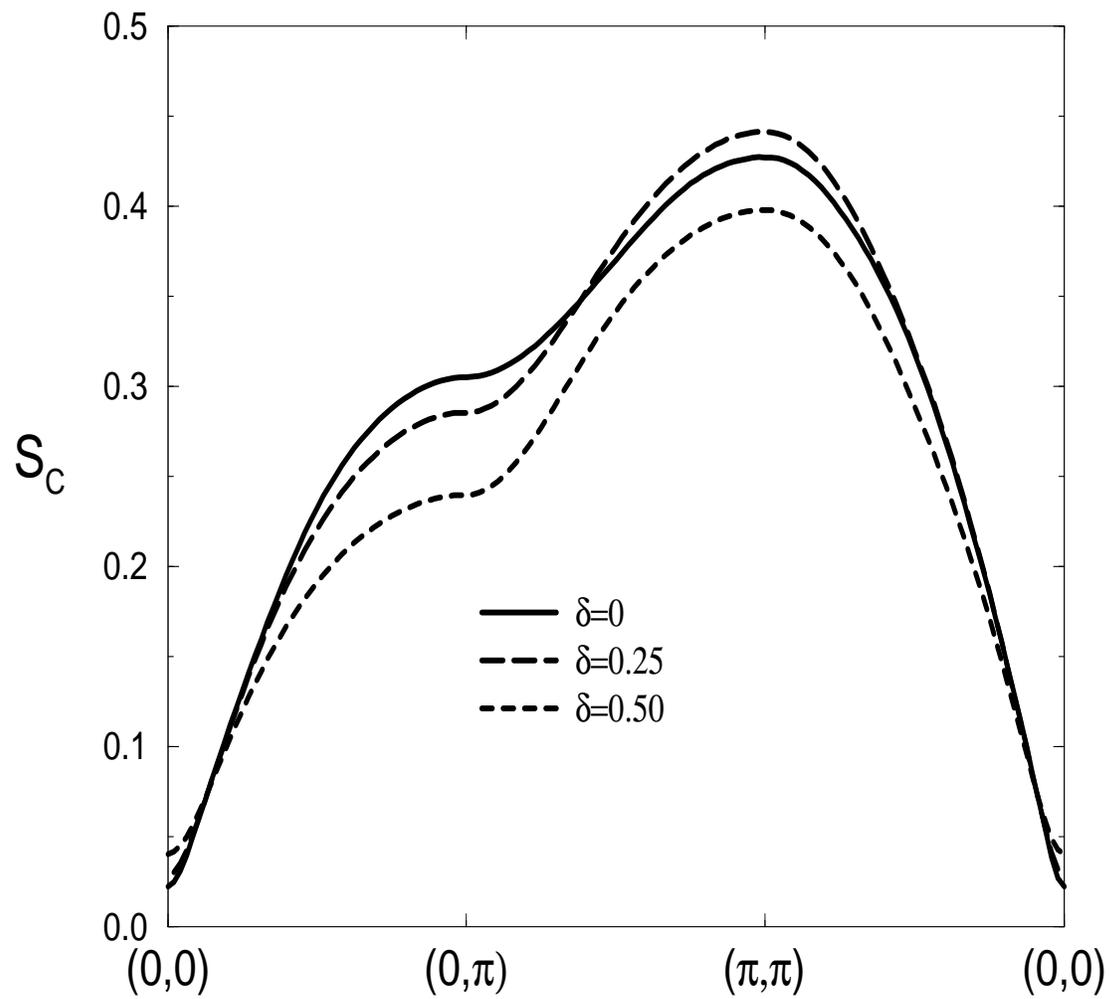,height=16cm,width=16cm}}
\caption{\label{fig5} Charge structure factor of the t-t'-U model for 
$U=4t$, $\beta=8$ and $t'=-0.47t$ for several values of $\delta$.}  
\end{figure}

We now proceed to the numerical results. We first calculate the
density dependence of the static (but $\vec q$-dependent) magnetic
susceptibility. In order to magnify the effect of $t'$, we perform the
calculation for $t'=-0.47t$. Had we chosen $t'=-0.5t$, then the van
Hove singularity would lie right at the lower band edge. For $U=4t$
and $\beta = 2$ we display the density-dependence of $\chi_s$ for
several $\vec q$-vectors in Fig. \ref{fig1}. For these parameters the
paramagnetic phase does not show magnetic instability. At a particular
doping the maximum of $\chi_s$ (in its $\vec q$-dependence) tells us
towards which phase an instability will develop. We checked
numerically that this really happens at lower temperature. In the
vicinity of half-filling $\chi_s$ is largest for the commensurate
vector $Q=(\pi,\pi)$. In the low-density range $\chi_s$ is maximal for
$q=0$. This range is very large and extends from $\delta \simeq 0.38$
to $\delta = 1$, $\delta$ being the hole doping. In this domain the 
fluctuations are predominantly ferromagnetic, because the system is
making use of the van Hove singularity to reduce its free
energy. Between these two 
regimes there is a small window
where the instability is towards an incommensurate phase with $\vec q$
along the diagonal of the Brillouin zone. We find that there is a
value of the doping, that we denote $\delta_0$, beyond which $\chi_s$
is largest for $q=0$. $\delta_0$ is seen to decrease with increasing
interaction. 
We note that the doping
$\delta_0$ at which $\chi_s$ is largest for $q=0$ decreases with
increasing interaction. For $U=0$ we found it to be $\delta_0 = 0.42$,
while for $U \geq 15t$ $\delta_0$ goes to zero. We thus obtain that,
for strong coupling, the paramagnetic phase is unstable towards
ferromagnetism over the entire doping range. This dependence of
$\delta_0$ on $U$ can be traced back to the $q$-dependence of the
effective interaction $A_{{\vec k}}$ entering Eq. (\ref{chis}), as
discussed below Eq. (\ref{blob}). It turns out that
the $q$-dependence is weak for weak coupling, and gets stronger with
increasing $U$. This plays a crucial role in assessing towards which
phase a magnetic instability may develop. We note that this effect is
neglected in the usual RPA and in the two-particle self-consistent
approach \cite{Vilk}. In those approaches all
what matters is the $q$-dependence of the bare susceptibility
$\chi_0$. We note that including $t'$ changes dramatically the phase
diagram as compared to the $t'=0$ case. In the latter case ferromagnetism
may only show up for very strong coupling ($U\geq 66t$) and in a
narrow doping region located around $\delta \simeq 15 \%$ \cite{Moller}. The
influence of $t'$ on the doping dependence of the uniform
susceptibility is displayed on Fig. \ref{fig2} for $U=4t$ and
Fig. \ref{fig3} for $U=20t$. For moderate  coupling decreasing $t'$
changes the monotonic behavior of $\chi_s(\delta)$ into a
non-monotonic one, which is typical of high-$T_c$ materials. The
height of the maximum increases with $t'$, and its location is shifted
towards higher doping. This behavior, as well as the location and the
height of the maximum, agree with the QMC data of
Veilleux {\it et al.} \cite{Veilleux}. We thus conclude that the
non-monotonic 
behavior of $\chi_s$ for moderate coupling mostly results from band-structure
effects. Experimentally the non-monotonic behavior of $\chi_S$ has
been observed in La$_{2-x}$Sr$_x$CuO$_4$ and the maximum is
reached for $x \simeq 0.25$ \cite{Torrance}; in YBa$_2$Cu$_3$O$_{7-x}$
$\chi_S$ only increases with increasing hole doping, and one
may assume that a maximum is reached for doping values that
cannot be reached experimentally. According to Hybertsen {\it et al.}
\cite{Hybertsen}, $t'=-0.16t$ is relevant to La$_{2-x}$Sr$_x$CuO$_4$,
in which case our approach yields the location of the maximum
of $\chi_S$ at $\delta \simeq 0.20$. However the dependence of
$\chi_S$ on $\delta$ is too weak to reproduce the experimental data.
In the strong coupling regime $\chi_s$ has a maximum in its
doping-dependence for $t'=0$ \cite{Doll}. Decreasing $t'$ results into an
enhancement of the maximum of $\chi_s$, and into a shift of it towards
larger doping. For large $t'$ its location coincides with the van
Hove singularity. Accordingly  this non-monotonic behavior of $\chi_s$
for strong coupling results from a combination of interaction and band
structure effects. We calculated $\chi_s(\vec q)$ for other values of
$\vec q$ too. For $t'=-0.47t$ it turned out that $\chi_s(\vec q=0)$ is
largest over the entire doping range. Thus raising up the interaction
leads to a dramatic widening of the ferromagnetic domain. This is in
agreement with the variational calculation of Pieri et
al. \cite{Pieri}, who investigated in more detail the low density
route to ferromagnetism due to M\"uller-Hartmann \cite{Muller}. \\
We now turn to the charge structure factor. The latter is calculated
according to Eq.(13) and Eq.(17) of Ref. \cite{Zimmermann}. We recall that the
comparison of the slave boson charge structure factor to existing QMC
data displays a quantitative agreement. Here we perform the calculation
for finite $t'$, for $U=4t$, $\beta =8$ and quarter-filling and
display the result on Fig. \ref{fig4}. As compared to the $t'=0$
result, decreasing $t'$ (i.e. $t'/t$ becoming increasingly
negative) substantially suppresses the charge structure factor,
especially around $(0,\pi)$, but also around $(\pi,\pi)$ for the
largest $t'$. This suppression results from a frustration of the
charge dynamics. We note that this suppression takes place in the
parameter regime where the tendency towards magnetism is
strongest. This leads us to propose that this is a general
situation: frustrated charge dynamics and magnetic instabilities are
occurring simultaneously in strongly correlated systems. This is well
known for the Hubbard model at half-filling where the charge dynamics
is so strongly frustrated that the system becomes insulating and
the physics is dominated by strong anti-ferromagnetic
fluctuations. Here the effect is less dramatic since the system
remains metallic, but clearly noticeable. In the opposite case of a
positive $t'$ (i.e. $t'/t >0$) the charge structure factor gets
essentially flat along the side of the Brillouin zone. It is
particularly enhanced in the vicinity $(0,\pi)$. This can be partly
understood using a local picture. Assuming that the charges can form
either a checkerboard or a stripe pattern, one sees that the number of
frustrated bonds is larger for the former pattern. This plays little
role if $t'$ is negative, since one hop along the diagonal costs
energy, but an increasingly important one when $t'$ is increasingly
positive. Clearly no such charge ordering occurs here, but tendencies
towards such patterns emerge. The frustration
effect is absent, and we indeed did not find any sign of a magnetic
instability. The correlation frustrated charge dynamics-enhanced
magnetic fluctuations can be traced back to the relationship between
the local susceptibilities and the density 
\begin{equation} \label{pauli}
\frac{1}{\beta} \sum_{{\vec k}, i\nu_n} \left( \chi_S({\vec k},
i\nu_n) + \chi_c({\vec k}, i\nu_n) \right) = 2 n \quad ,
\end{equation} 
which follows from the Pauli principle. Indeed if the local
charge susceptibility is reduced, the magnetic one is enhanced, and
{\it vice versa}.
In Fig. \ref{fig5} we display the density dependence of
the charge structure factor, for $U=4t, \beta=8$ and
$t'=-0.47t$. Under an increase of the density, $S_c$ first goes up,
until $\delta \simeq 0.25$ and then goes down while going closer to
half-filling. In weak coupling one would expect $S_c$ to decrease upon
doping, but the opposite behavior holds in a dense strongly
correlated system. 

In summary we studied the charge and magnetic
properties of the $2$-dimensional $t-t'-U$ model. We found that a negative $t'$
has a strong influence on the phase diagram, a very large portion of
it being dominated by strong ferromagnetic fluctuations. We also
showed that frustrated charge dynamics and strong magnetic
fluctuations occur simultaneously. We found a tendency toward striped
phases only for large positive $t'$.

\section{Acknowledgments}
We thank P. W\"olfle for valuable comments and his steady
encouragement. We are grateful to T. Kopp, H. Kroha, H. Beck and M
Capezzali for
useful discussions. R. F. is grateful for the warm hospitality at the
Institut f\"ur Theorie der Kondensierten Materie of Karlsruhe University, where
part of this work has been done.
This work has been supported by the Deutsche Forschungsgemeinschaft 
through Sonderforschungsbereich 195. One of
us (RF) is grateful to the Fonds National Suisse de la Recherche Scientifique
for financial support.


\end{document}